\documentclass[aps,pra,twocolumn,showpacs]{revtex4}
\usepackage{graphicx}
\usepackage{graphics}
\usepackage{epsf}
\usepackage{epstopdf}
\usepackage{psfrag}
\usepackage[T1]{fontenc} 
\usepackage{epsfig,latexsym}
\usepackage{color}
\usepackage{amsmath}
\usepackage{amssymb}
\usepackage{amsfonts}
\usepackage{indentfirst}
\usepackage{ae}
\usepackage{float}

\newcommand{\be}{\begin{equation}}
\newcommand{\bq}{\begin{eqnarray}}
\newcommand{\eq}{\end{eqnarray}}

\begin{document}
\title{Exotic looped trajectories via quantum marking}

\author{J. G. G. de Oliveira Jr. $^{1}$}\email[]{jgojunior@uesc.br}
\author{Gustavo de Souza$^{2}$}
\author{L. A. Cabral$^{3}$}
\author{I. G. da Paz$^{4}$}
\author{Marcos Sampaio$^{5}$}

\affiliation{$^1$ Departamento de Ci\^{e}ncias Exatas e Tecnol\'{o}gicas --
Universidade Estadual de Santa Cruz, \\
45.662--900, Ilh\'{e}us -- BA, Brazil}

\affiliation{$^{2}$ Universidade Federal de Ouro Preto -
Departamento de Matem\'{a}tica - ICEB Campus Morro do Cruzeiro, s/n,
35.400 -- 000, Ouro Preto MG - Brazil}

\affiliation{$^3$ Curso de F\'{\i}sica, Universidade Federal do
Tocantins, Caixa Postal 132, CEP 77804-970, Aragua\'{\i}na, TO,
Brazil}

\affiliation{$^4$ Departamento de F\'{\i}sica, Universidade Federal
do Piau\'{\i}, Campus Ministro Petr\^{o}nio Portela, CEP 64049-550,
Teresina, PI, Brazil}

\affiliation{$^{5}$ Departamento de F\'{\i}sica, Instituto de
Ci\^{e}ncias Exatas, Universidade Federal de Minas Gerais, Caixa
Postal 702, CEP 30161-970, Belo Horizonte, Minas Gerais, Brazil}

\begin{abstract}
We provide an analytical and theoretical study of exotic looped trajectories (ELTs) in a double-slit interferometer with quantum marking.~We use an excited Rydberg-like atom and which-way detectors such as superconducting cavities, just as in the Scully-Englert-Walther interferometer.~We indicate appropriate conditions on the atomic beam or superconducting cavities so that we determine an interference pattern and fringe visibility exclusive from the ELTs.~We quantitatively describe our results for Rubidium atoms and propose this framework as an alternative scheme to the double-slit experiment modified to interfere only these exotic trajectories.
\end{abstract}


\maketitle

\section{Introduction}
\label{sec:intro}

\par Exotic looped trajectories (ELT) in multi-slit interferometry has
emerged as an interesting arena to test foundations of quantum
mechanics both theoretically and experimentally. More specifically,
one may exploit the validity of the superposition principle and the
Born rule \cite{Born} to compute probabilities from wave functions
which connects quantum theory with experiment. Born rule implies
that in multi-slit interferometry the interference pattern reduces to
a sum of terms denoting the interference between pairs of
wavefunctions. Any deviation from this construction in the intensity
on the screen would possibly indicate a violation of Born's rule.

\vspace{0.2cm}
\par For double slits $A$ and $B$, the probability of detection at a point $x$ on the screen is given, according to the Born rule, by
\begin{equation}\label{eq:P_ij}
P_{AB} = P_A + P_B + (\psi_A^*\psi_B  + h.c.) ,
\end{equation}
\noindent with $P_{i} =|\psi_{i} |^2 $, $i=A,B$. On the other hand, for triple slits $A$, $B$ and $C$, the interference pattern observed 
at $x$ reads
\begin{equation}
P_{ABC} = P_{A\,B} + P_{B\,C} + P_{A\,C} - P_A - P_B - P_C \, ,
\end{equation}
\noindent with $P_{ij}$ given by eq.~(\ref{eq:P_ij}) for $ij=AB, BC, AC$,
and $P_{i} =|\psi_{i} |^2 $, $i=A,B,C$. The wave functions $\psi_{i}$, with $i=A,B,C$, represent the probability amplitude of the particle to arrive at the point $x$ on the detection screen having emerged from a source placed before the grating (say, centered right at the middle slit) and passed through slit $(A,B,C)$. A deviation from the ``naive" Born rule above through a tiny parameter, called the Sorkin Parameter \cite{Sorkin, Sinha1}
\begin{equation}
\kappa_{S} = P_{ABC} - P_{A\,B} - P_{B\,C} - P_{A\,C} + P_A + P_B + P_C\, ,
\end{equation}
\noindent can be attributed to exotic trajectories which also come in pairs. With support on the Feynman path integral prescription, one should also include looped paths around the slits (paths that go backwards in space variables), which also appear in pairs. Taking into account the leading order contributions leads to
\begin{eqnarray}
\kappa_{S}^{\mathrm{et}} = &\kappa_S - P_{A\,ABA} - P_{A\,BAB} - P_{B\,ABA} \nonumber \\
&- P_{B\,BAB} - P_{ABA\,BAB}\, .
\end{eqnarray}
\noindent In this expression, $ABA$ (resp., $BAB$) stands for clockwise (resp., counterclockwise) loops, and the superscript $\mathrm{et}$ in $\kappa $ stands for ``exotic trajectories''. The last term in it contributes less, as well as loops
that wind repeatedly through the slits \cite{footnote1}.

\vspace{0.2cm}
\par To the best of our knowledge the first theoretical treatment of the contribution of the so called ``non-classical" or ``exotic" paths in matter wave double-slit interferometry was by Yabuki in \cite{Yabuki}, in which nonlinear interference terms were used to estimate deviations from the superposition principle based on the Feynman path integral approach \cite{FeynmanHibbs} to include looped paths around the slits. This was later discussed in a similar way by Sorkin in in \cite{Sorkin}, with an eye towards experiments. In Sorkin's work, higher-order phenomena were added to the usual prescription of interference when three or more paths interfered. It was only more recently that quantification of the contribution from exotic paths in interference experiments for triple-slits was proposed \cite{Sinha1,Raedt,Sinha2,Sinha3}.

\vspace{0.2cm}
\par In 2016 the first experimental observation of exotic paths in triple-slit interference with light was presented in \cite{Boyd,USinha}. In the experiment, exotic trajectory contributions to interference fringes were enhanced by controlling the strength and spatial distribution of electromagnetic near fields at the vicinity of the slits by the manipulation of surface plasmons. Some current experimental investigations have also searched for deviations from quantum theory by looking at higher-order interference \cite{Jin}. These studies are putting stringent constraints on potential failures of Born's rule. Other proposals involve passing a particle through a physical barrier with multiple slits and comparing the interference patterns formed on a screen behind the barrier when different subsets of the slits are closed \cite{Barnum}. Additionally, interest in consequences of the existence of higher-order interference has increased recently because of the possibility of applications in different physical systems \cite{Lee}.

\vspace{0.2cm}
\par The theoretical analysis that support the experiments is based on path integrals in the presence of slits, with different weights for classical and exotic trajectories. The propagator is written as $$K(\vec{r}_1,\vec{r}_2)=\int {\cal{D}} [\vec{x}(s)] \exp[i k \int ds],$$ where $s$ is the contour length along $\vec{x}(s)$, the classical limit being $k\rightarrow \infty$ with paths near the straight line linking $\vec{r}_1$ to $\vec{r}_2$ contributing more
by the stationary phase approximation (to leading order). Paths away
from the classical path contribute with a rapidly oscillating phase.

\vspace{0.2cm}
\par Triple-slit interferometers have become an useful testing ground
for checking Born's rule in quantum mechanics \cite{Sorkin, Sinha1, Park, Niestegge}. Experiments with triple slits proposed in \cite{Sinha3} using matter waves or low frequency photons were analytically described and allowed for an estimate of an upper bound on the Sorkin factor $|\kappa_{max}| \approx 0.003 \lambda^{3/2}/(d^{1/2} w)$, in which $\lambda$ is the wavelength, $d$ is the center-to-center distance between the slits and $w$ is the slit width, confirming that $\kappa$ is rather sensitive to the experimental setup. In \cite{Sinha2}, the exotic trajectories contribution for a triple-slit matter wave diffraction was evaluated using the non-relativistic Feynman path integral approach with a free propagator given by $K(\vec{r},\vec{r}^{\prime})=\frac{k}{2\pi i}\frac{1}{|\vec{r}-\vec{r}^{\prime}|}\;e^{ik|\vec{r}-\vec{r}^{\prime}|}$. This satisfies the Helmholtz equation away from $\vec{r}=\vec{r}^{\prime}$ and the Fresnel-Huygens principle. In the Fraunhofer regime, it leads to integrals which are evaluated numerically using the stationary phase approximation. As a result, it follows that $\kappa\approx10^{-8}$ for electron waves.

\vspace{0.2cm}
\par Evidently, the effect of exotic paths can also be computed in double-slit setups, but no experimental deviation from $\kappa = 0$ in this type of arrangement has been reported as yet. Although the first results for exotic path contributions were obtained within the triple-slit setup, double-slit apparatuses can be more appropriate to demonstrate the effect, since for the triple-slit one can actually observe effects of high-order interference instead of exotic paths \cite{Jin}.~Recently, some of us proposed to observe effects of exotic looped trajectories in double-slit interference by relating the Sorkin parameter to the relative intensity and the fringe visibility \cite{Paz3}.~In the present paper, a different approach is presented:~withdraw the contribution of classical paths through quantum marking and display the interference pattern of exotic looped trajectories only.~We can realize this by the resonant interaction of Rydberg atoms with single field modes of high-finesse cavities positioned in the slit apertures.~Our procedure is based on quantum marking, with the microwave cavities used as which-way detectors.~We determine the fringe visibility of the ELT, and show how it can be increased by appropriate measurement of the atomic beam or the superconducting cavities.~We quantitatively describe our results for Rubidium atoms and propose this framework as an alternative scheme to measure exotic looped paths and deviations from the standard superposition principle.~Furthermore, this attainment can be further important in possible applications involving interference of only the looped trajectories.

\vspace{0.2cm}
\par This paper is organized as follows. After a discussion of the general formalism for Born's rule (Section \ref{section:um}) and which-way detectors (Section \ref{section:dois}) in two-way interferometry, we specialize in Section \ref{section:2Slit_ELT} to atoms interacting with a single field mode in a microwave cavity. We present a two-slit interferometer where the contribution of classical paths can be withdrawn through quantum marking. Next, we analyze in Section \ref{section:results} the fringe visibility of the ELT and study the interference pattern of these exotic paths for Rubidium atom waves. Finally, a discussion of the results is made in Section \ref{section:discussion} together with a few concluding remarks.

\section{Born's rule for wave functions}
\label{section:um}

\par Born's rule \cite{Born}, formulated in 1926, predicts that quantum  interference in a multipath interferometer occurs from pairs of paths, a direct consequence of the probability of an outcome $x$ for a system described by a wavefunction $\psi(x)$ being the absolute square of this function.

\vspace{0.2cm}
\par
\begin{figure}[h]
\centering
\includegraphics[scale=0.3,angle=00]{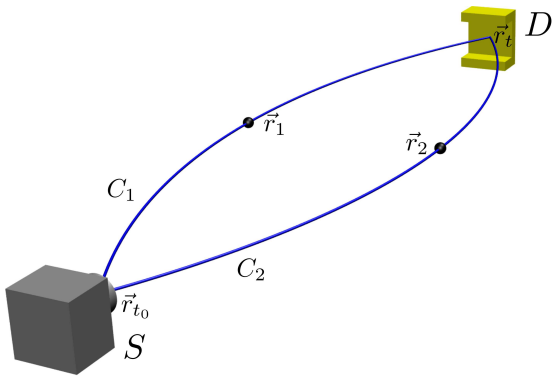}
\caption{\footnotesize Free evolution from $\mathbf{r}_{t_{0}}$ to
$\mathbf{r}_{t_{}}$ during $\Delta t$. In the course of evolution, the
particle passes through either $\mathbf{r}_{1}$ or $\mathbf{r}_{2}$.
$C_1$ and $C_2$ are possible paths before detection by $D$.}
\label{fig1}
\end{figure}

 In the classical formulation of Born's rule, a particle described by an initial quantum state $|\psi_{t_0}\rangle$, produced in a small region around $\mathbf{r}_{t_{0}}$, evolves freely during $\Delta t=t-t_0 $ before being detected by a device $D$ at $\mathbf{r}_{t_{}}$. This happens after the particle has passed through either $\mathbf{r}_{1}$ (trajectory $C_1$) or $\mathbf{r}_{2}$ (trajectory $C_2$) -- see Figure \ref{fig1}. Then,  Born's rule is the statement that the state of the particle at $\mathbf{r}_{t}$, right before its detection by $D$, is
\begin{equation}\label{eq1.1}
|\psi_t\rangle=\mathcal{N}\bigl(a_1|\psi_t\rangle_{C_1}+a_2|\psi_t\rangle_{C_2}\bigr),
\end{equation}
\noindent where $\mathcal{N}$ is a normalization constant and
$|\psi_t\rangle_{C_i}$ the evolution from $\mathbf{r}_{t_0}$ until
$\mathbf{r}_{t}$ through $\mathbf{r}_{i}$, with associated probability amplitudes $a_i$,  $i=1,2$.

\vspace{0.2cm}
\par The state given by equation (\ref{eq1.1}) is a quantum superposition which represents a generic two-way interferometer output state. At the detector $D$ the intensity reads
\begin{eqnarray}
\label{eq1.2}
&& I(\mathbf{r}_t) = \langle\mathbf{r}_t|\psi_t\rangle\langle\psi_t|\mathbf{r}_t\rangle = \mathcal{N}^2\times \\
&&\Bigl[I_1(\mathbf{r}_t)+I_2(\mathbf{r}_t) +\bigl(
a_1 a_{2}^{*}\langle\mathbf{r}_t|\psi_t\rangle_{C_1}\,_{C_2}\langle\psi_t|
\mathbf{r}_t\rangle + \mathrm{h.c.}\bigr)\Bigr] \nonumber,
\end{eqnarray}
\noindent in which $I_i(\mathbf{r}_t)=|a_i|^2\langle\mathbf{r}_t|\psi_t\rangle_{C_i}
\,_{C_i}\langle\psi_t|\mathbf{r}_t\rangle$,  $i=1,2$. It is useful to define the visibility (wave character) through Eq. (\ref{eq1.2}),
\begin{equation}
V=\dfrac{2|a_1 a_{2}\langle\mathbf{r}_t|\psi_t\rangle_{C_1}\,_{C_2}\langle\psi_t|
    \mathbf{r}_t\rangle |}{I_1(\mathbf{r}_t)+I_2(\mathbf{r}_t)}\label{V},
\end{equation}
\noindent as well as the predictability (particle character):
\begin{equation}
P=\dfrac{|I_1(\mathbf{r}_t)-I_2(\mathbf{r}_t)|}{I_1(\mathbf{r}_t)
    +I_2(\mathbf{r}_t)}\label{P}.
\end{equation}
\noindent These quantities satisfy a wave particle duality relation $P^2+V^2=1$.
%
\section{Which-way detectors}
\label{section:dois}

\par
\begin{figure}[h]
    \centering
    \includegraphics[scale=0.3,angle=00]{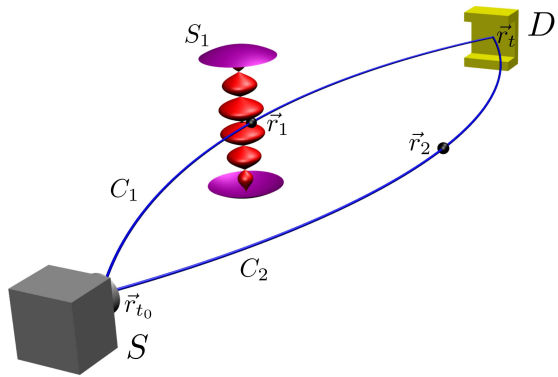}
    \caption{\footnotesize Quantum particle in free evolution from
        $\mathbf{r}_{t_{0}}$ to $\mathbf{r}_{t_{}}$ over a lapse of time
        $\Delta t$.  $C_1$ e $C_2$ are possible paths before detection in
        $D$. At $\mathbf{r}_{1}$, we place a which-way detector $S_1$ which
        has two orthonormal levels $\{|S_{-}^{(1)}\rangle,
        |S_{+}^{(1)}\rangle\}$. Initially $S_1$ is in the state
        $|S_{-}^{(1)}\rangle$. It will change to $|S_{+}^{(1)}\rangle$ only
        if the particle passed by $\mathbf{r}_1$ following $C_1$.}
    \label{fig2}
\end{figure}

 Assume that a which-way detector $S_1$ has been placed at $\mathbf{r}_1$, as depicted in Figure \ref{fig2}.~The bipartite system particle/which-way detector is considered to undergo an instantaneous interaction which does not significantly affect the particle's center-of-mass momentum.~$S_1$ is in the initial state $|S_{-}^{(1)}\rangle$ and will collapse to the state $|S_{+}^{(1)}\rangle $ if the particle follows path $C_1$.~Thus, the resulting state is an entangled state of the form
\begin{equation}\label{eq2.1}
|\Psi_t\rangle=\mathcal{N}\bigl(a_1|\psi_t\rangle_{C_1}|S_{+}^{(1)}\rangle+a_2|\psi_t
\rangle_{C_2}|S_{-}^{(1)}\rangle\bigr).
\end{equation}
\noindent Partial trace over the detector variables in the state  $|\Psi_t\rangle\langle\Psi_t|$ yields the following reduced state for the particle:
\begin{equation}\label{eq2.2}
\rho_t=\mathcal{N}^{2}\Bigl[|a_1|^2|\psi_t\rangle_{C_1}\,_{C_1}\langle\psi_t| +
|a_2|^2|\psi_t\rangle_{C_2}\,_{C_2}\langle\psi_t|\Bigr].
\end{equation}
\noindent In this case, the intensity at the detector $D$, $\langle\mathbf{r}|\rho_t|\mathbf{r}\rangle$, is
\begin{equation}\label{eq2.3}
\tilde{I}(\mathbf{r}_t) =
\mathcal{N}^2\Bigl[I_1(\mathbf{r}_t)+I_2(\mathbf{r}_t)\Bigr],
\end{equation}
\noindent which shows that interference terms are absent in (\ref{eq2.3}) (zero visibility). Should we impose that $I_1(\mathbf{r}_t)=I_2(\mathbf{r}_t)$, then the predictability would also vanish, yielding $P^2+V^2=0$. The interference fringes would have been washed out due to the entanglement between the particle and $S_1$, which makes which-way information to become available in the device.

\vspace{0.2cm}
\par Now, if a projective measurement with respect to a judiciously chosen basis for $S_1$ is made {\it{ before}} the particle is detected in $D$, the which-way information would be erased (quantum eraser). In order to see this, let us define the orthonormal basis $\{|+^{(1)}\rangle,|-^{(1)}\rangle\}$ for $S_1$ state space as
\begin{eqnarray}
|+^{(1)}\rangle &=& \dfrac{|S_{+}^{(1)}\rangle+|S_{-}^{(1)}\rangle}{\sqrt{2}}\label{m}\\
|-^{(1)}\rangle &=& \dfrac{|S_{+}^{(1)}\rangle-|S_{-}^{(1)}\rangle}{\sqrt{2}}\label{n}.
\end{eqnarray}
\noindent In terms of this basis, the state represented by equation (\ref{eq2.1}) translates into
\bq
|\Psi_t\rangle &=& \frac{\mathcal{N}}{\sqrt{2}}\Big[\bigl(a_1|\psi_t\rangle_{C_1}+a_2|\psi_t\rangle_{C_2}\bigr)|{+}^{(1)}\rangle \nonumber \\ &+&
\bigl(a_1|\psi_t\rangle_{C_1}-a_2|\psi_t\rangle_{C_2}\bigr)|{-}^{(1)}\rangle \Big].
\label{eq21}
\eq
\noindent After performing a measurement of $S_1$ relative to the basis $\{|+^{(1)}\rangle,|-^{(1)}\rangle\}$, the composite particle/which-way detector state reduces to
\begin{equation}\label{pm1}
|{\psi}_{t}^{\pm}\rangle|\pm^{(1)}\rangle = \mathcal{N}
\bigl(a_1|\psi_t\rangle_{C_1}\pm
a_2|\psi_t\rangle_{C_2}\bigr)|\pm^{(1)}\rangle \, .
\end{equation}
\noindent The particle will be detected in $D$ with intensity (after tracing
out the which-way detector degrees of freedom)
\begin{eqnarray}
&& I(\mathbf{r}_t) ^{\pm} = \langle\mathbf{r}_t|\psi_{t}^{\pm}\rangle
\langle\psi_{t}^{\pm}|\mathbf{r}_t\rangle = \mathcal{N}^{\,2} \times \\
 && \hskip-0.4cm \Bigl[I_1(\mathbf{r}_t)+I_2(\mathbf{r}_t) \pm\Bigl(
a_1 a_{2}^{*}\langle\mathbf{r}_t|\psi_t\rangle_{C_1}\,_{C_2}\langle\psi_t|
\mathbf{r}_t\rangle + \mathrm{h.c.}\Bigr)\Bigr] \nonumber\label{Ipm}.
\end{eqnarray}
\noindent Clearly, if we measure the which-way detector in the state $|+^{(1)}\rangle$ we recover the fringes as in equation (\ref{eq1.2}) (quantum erasure); otherwise, if we get $|-^{(1)}\rangle$, we obtain anti-fringes. Moreover: (i) equations (\ref{V}) and (\ref{P}) hold, and hence $P^2+V^2=1$, and (ii) if $I_1(\mathbf{r}_t)=I_2(\mathbf{r}_t)$, then $P=0$ and $V=1$.

\vspace{0.2cm}
\par
\begin{figure}[h]
\centering
\includegraphics[scale=0.3,angle=00]{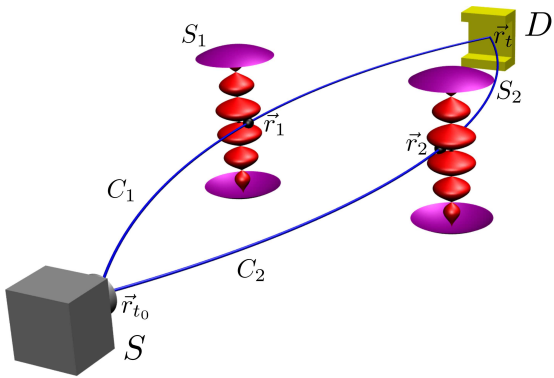}
    \caption{\footnotesize Free evolution of a quantum particle from $\mathbf{r}_{t_{0}}$ to
    $\mathbf{r}_{t_{}}$ during $\Delta t$, passing by $\mathbf{r}_{1}$ or $\mathbf{r}_{2}$. $C_1$ and $C_2$ are possible trajectories before detection in $D$. In $\mathbf{r}_{1}$ (resp., $\mathbf{r}_{2}$), a which-way detector $S_1$ (resp., $S_2$) possesses two orthonormal states $\{|S_{-}^{(1)}\rangle, |S_{+}^{(1)}\rangle\}$ (resp., $\{|S_{-}^{(2)}\rangle,
    |S_{+}^{(2)}\rangle\}$). Initially at the state
    $|S_{-}^{(1)},S_{-}^{(2)}\rangle $, it will assume the state $|S_{+}^{(1)},S_{-}^{(2)}\rangle$ (resp., $|S_{-}^{(1)}, S_{+}^{(2)}\rangle$) in case the particle follows $C_1$ (resp., $C_2$).}
    \label{fig3}
\end{figure}

 One could also set up a second which-way marker at $\mathbf{r}_2$ -- namely $S_2$, see Figure \ref{fig3} ahead. Again, assuming the particle in the initial state $|\psi_{t_0}\rangle$ and leaving from $\mathbf{r}_{t_0}$ to be detected in $D$, the final state of the tripartite system at $\mathbf{r}_t$ just before arriving at the detector $D$ will in this case
\begin{equation}\label{eq2.9}
|\Phi_t\rangle=\mathcal{N}
\bigl(a_1|\psi_t\rangle_{C_1}|S_{+}^{(1)}\rangle|S_{-}^{(2)}\rangle+a_2|\psi_t
\rangle_{C_2}|S_{-}^{(1)}|S_{+}^{(2)}\rangle\bigr)\, ,
\end{equation}
\noindent where $S_1$ ($S_2$) is in the initial state $|S_{-}^{(1)}\rangle$ 
($|S_{-}^{(2)}\rangle$) and will collapse to the state $|S_{+}^{(1)}\rangle$ ($|S_{+}^{(2)}\rangle$) if the particle follows by 
$C_1$ ($C_2$), respectively.~The which-way information in $S_1$ and $S_2$ gives, after tracing the detectors degrees of freedom, a state just as in (\ref{eq2.2}).~The intensity measured by $D$ is as in equation (\ref{eq2.3}).
\par Finally, assume that in the case when two which-way markers are available one performs Bell measurements -- namely, joint measurements in the which-way detectors whose result is a projection into a Bell state --, also before detection in $D$. For that purpose, let us rewrite 
$\{|S_{+}^{(1)},S_{+}^{(2)}\rangle,\,|S_{+}^{(1)},S_{-}^{(2)}\rangle,\,
|S_{-}^{(1)},S_{+}^{(2)}\rangle,\,|S_{-}^{(1)},S_{-}^{(2)}\rangle\}$ in the Bell basis $\{|\psi_{\pm}\rangle,|\phi_{\pm}\rangle\}$:
\begin{eqnarray}
|\psi_{+}\rangle=\dfrac{|S_{+}^{(1)},S_{-}^{(2)}\rangle+|S_{-}^{(1)},S_{+}^{(2
)}\rangle}{\sqrt{2}}\label{psim}\\
|\psi_{-}\rangle=\dfrac{|S_{+}^{(1)},S_{-}^{(2)}\rangle-|S_{-}^{(1)},S_{+}^{(2
)}\rangle}{\sqrt{2}}\label{psin}\\
|\phi_{+}\rangle=\dfrac{|S_{+}^{(1)},S_{+}^{(2)}\rangle+|S_{-}^{(1)},S_{-}^{(2
)}\rangle}{\sqrt{2}}\label{phim}\\
|\phi_{-}\rangle=\dfrac{|S_{+}^{(1)},S_{+}^{(2)}\rangle-|S_{-}^{(1)},S_{-}^{(2
)}\rangle}{\sqrt{2}}\label{phin}.
\end{eqnarray}
\noindent In such a basis, the state displayed in (\ref{eq2.9}) becomes
\bq
\label{eq2.14}
|\Phi_t\rangle &=& \frac{\mathcal{N}}{\sqrt{2}}\Bigl[\bigl(a_1|\psi_t\rangle_{C_1}+a_2|\psi_t\rangle_{C_2}\bigr)|\psi_{+}\rangle
\nonumber \\ &+&
\bigl(a_1|\psi_t\rangle_{C_1}-a_2|\psi_t\rangle_{C_2}\bigr)|\psi_{-}\rangle\Bigr].
\eq
\noindent Notice that the states represented by equations (\ref{eq21}) and
(\ref{eq2.14}) are formally identical substituting
$|\pm^{(1)}\rangle $ with $|\psi_\pm\rangle$. Suppose that the measurements proceed just as we described for $S_1$, save that now we perform
Bell measurements in the subsystems $S_1$ and $S_2$. We
have two detectors and four Bell states, two of the measurements
yielding zero $|\langle\phi_{\pm}|\Phi_t\rangle|^2=0$, whilst
the other two $|\langle\psi_{\pm}|\Phi_t\rangle|^2=1/2 $ are
equiprobable. Then, after a joint Bell measurement in $S_1$ and $S_2$, and immediately before detection in $D$, the overall state of the
system would be
\begin{equation}
|{\chi }_t^{\pm}\rangle = \mathcal{N}\bigl(a_1|\psi_t\rangle_{C_1}
\pm a_2|\psi_t\rangle_{C_2}\bigr)|\psi_{\pm}\rangle.
\end{equation}
\noindent The intensity detected in $D$ in this case can be computed from
$\langle\mathbf{r}_t|\chi _{t}^{\pm}\rangle\langle
\chi _{t}^{\pm}|\mathbf{r}_t\rangle $ to yield the same result as in
(\ref{Ipm}).~Just as in the case where one which-way detector was
involved, we have fringes and anti-fringes with equal probability.~Bell measurements in $S_1$ and $S_2$ erase the information about which path the particle has followed, and thus quantum coherence is restored.~Of course the visibility $V=1$ when $I_1(\mathbf{r}_t)=I_2(\mathbf{r}_t)$.
%
\section{Double slit interference of exotic looped trajectories}
\label{section:2Slit_ELT}

\par Specializing the above to a QED cavity setting ({\it e.\,g.}, see \cite{livro} and references therein), we are now in position to present a two-slit interferometer where the contribution of classical paths is withdrawn through quantum marking, but where we can still observe the interference pattern of only the exotic looped trajectories.

\vspace{0.2cm}
\par
\begin{figure}[h]
    \centering
   \includegraphics[scale=0.27,angle=00]{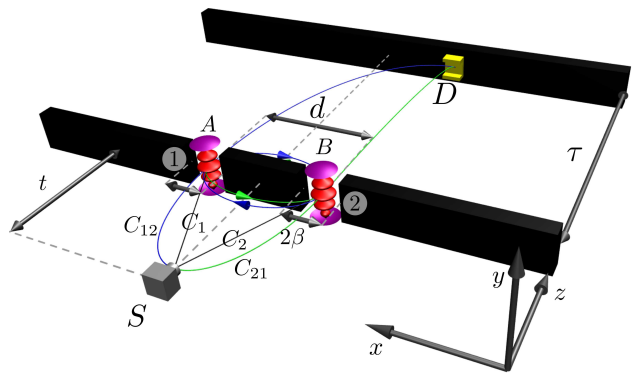}
    \caption{\footnotesize Sketch of the double-slit experiment modified to interfere only exotic looped trajectories.~$S$ is a source that sends atoms and is located on the axis of symmetry of the experiment.~In the source $S$, two-level atoms of transition frequency $\omega_{ge}$ are produced with the internal state in $|e\rangle$, propagate for a time $t$ before reaching the double-slit, and then for a time $\tau$ from the double-slit to the detector $D$.~The slit apertures are taken to be Gaussian, of width $\beta$, and separated by a distance $d$. Cavity $A$ (resp. $B$) is positioned in slit $1$ (resp., $2$).~These cavities are initially in the vacuum, are identical, and have the same frequency $\omega_{c}$. The wavefunction for the center-of-mass of the atom is Gaussian with transverse width $\sigma_{0}$. From the source to the detector the atoms can follow one of the paths $C_1$ and $C_2$ (non-exotic paths) or $C_{12}$ and $C_{21}$ (looped trajectories or exotic paths). By resonant interaction between the atoms and the field inside the cavities ($\omega_{ge}=\omega_{c}$), it is possible to mark the paths $C_{1}$ and $C_{2}$ and observe the interference of the paths $C_{12}$ and $C_{21}$.}
    \label{fig4}
\end{figure}

 We consider for this purpose the double slit interferometer depicted in Figure \ref{fig4}, where $S$ is the source, $t$ is the propagation time before reaching the double-slit, $\tau$ is the propagation time from the double-slit to the detector $D$, and $d$ is the interslit separation. The paths $C_1$, $C_2$ are non-exotic paths and the paths $C_{12}$ (blue or clockwise loop) and $C_{21}$ (green or counterclockwise loop) are looped trajectories, or exotic paths. Besides the usual contributions from paths $C_1$ and $C_2$ we shall consider two leading order exotic looped path contributions to the intensity at the detections screen, $C_{12}$ and $C_{21}$. From the point of view of Feynman's sum over paths to express the
quantum propagation amplitude, such paths correspond to trajectories that go forward and backward in space selected by the (two-)slitted barrier. Trajectories at large distances from the classical trajectory contribute less to the propagation amplitude as they are modulated by a rapidly oscillating phase in the Feynman path integral \cite{FeynmanHibbs}.

\vspace{0.2cm}
\par In this interferometer, we take the two-slitted potential barrier to be represented by a Gaussian function centered at the slit with an effective width corresponding to the size of the slit aperture.~We consider 
circular Rydberg atoms \cite{atomo} with two-levels, i.e.,
$\left| g \right\rangle $ (ground state) and $\left|e\right\rangle $ (excited state) (separated by the energy difference $E_{e}-E_{g}=\hbar\omega_{ge}$), produced in the source
$S$ and sent one-by-one to the double slit where there are QED
cavities $A$ (at slit $1$) and $B$ (at slit $2$), see Figure \ref{fig4}.~We will take $A$ and $B$ as being open single mode cavities in the Fabry--Perot configuration and the effects of losses will be disregarded. The atoms interact resonantly with a mode in the cavity and the Hamiltonian of interaction will be
\begin{equation}\label{jc}
H_{JC}=-i\hbar\dfrac{\Omega_{0}}{2}(\hat{a}\hat{\sigma}_{+}-\hat{a}^\dagger\hat{\sigma}_{-}),
\end{equation}
\noindent which is the Jaynes--Cummings model \cite{interacao,livro}  in the rotating wave approximation, where $\Omega_0$ is the vacuum Rabi 
frequency, $\hat{a}$ ($\hat{a}^{\dagger}$) is the bosonic operator 
that destroys (creates) field's excitations in the cavity, and
$\hat{\sigma}_{-}$ ($\hat{\sigma}_{+}$)
is the fermionic operator that lowers (raises) the internal states of the 
atom, respectively.~Thus, when the atom arrives at the detector we can know about which slit it came in the case when it undergoes non-exotic propagation.~This setup consists of a classical proposal by Scully, Englert and Walther in Ref. \cite{Scully_Englert_Walther}. As discussed there, we can assume that the center-of-mass position wavefunction of the atoms are not disturbed by the interaction with the microwave field when
traversing the cavities.~In particular, this means that the atomic
center-of-mass state's evolution between locations inside the interferometer can be safely described by free particle propagators.

\vspace{0.2cm}
\par Present-day technology allows implementation of this setup with Rubidium atoms and high-Q superconducting cavities that can trap the microwave photons \cite{livro}.~After a photon has been trapped in such a cavity, it can be removed by passing through it another Rubidium atom in its ground state.~The photon will be absorbed also with unit probability.~Thus, if the atom makes an exotic looped propagation passing two times through the same slit, the information stored in the first passage is withdrawn in the second passage, and when it arrives in the detector we loose information about which slit it passed.~As we will see, this is the essential ingredient that allows us to select only the exotic path contributions to the interference pattern at the detection screen (see Figure \ref{fig4}).

\vspace{0.2cm}
\par For the sake of simplicity we consider that the spatial part of the atomic state produced in the source $S$ is a Gaussian function of transverse width $\sigma_{0}$. We also consider that motion along the $z$-direction is classical, and that the quantum effects are relevant only in the $x$-direction as previously treated in \cite{Paz4}. This allows for a complete analytical computation of the free propagation, global and local axial phases comprised.
%

\vspace{0.2cm}
\par The relevant physical degrees of freedom to be taken into account in
this setup are the atomic center-of-mass motional state (i.e., the
quanton's positional state inside the interferometer), the internal
excitation state of the atom, and the state of the photon field in
the superconducting cavities. The associate quantum state space in
each case will be denoted, respectively, by
$\mathcal{H}_{\mathrm{c.m.}}$, $\mathcal{H}_{\mathrm{i.s.}}$, and
$\mathcal{H}_{\mathrm{cav}}=\mathcal{H}_A \otimes \mathcal{H}_B$,
where $\mathcal{H}_A $ (resp., $\mathcal{H}_B $) denotes the one
photon sector of the Fock space for the allowed modes for cavity $A$
( $B$). The full composite system state space is
$\mathcal{H}=\mathcal{H}_{\mathrm{c.m.}}\otimes\mathcal{H}_{\mathrm{i.s.}}
\otimes\mathcal{H}_{\mathrm{cav}}$.

\vspace{0.2cm}
\par Now, if only non-exotic trajectories are taken into account, then 
we could write the quantum state of the full composite system at the 
interferometer screen immediately before detection as
\begin{equation}\label{eq:St_Sc_NonExotic}
|\Psi \rangle = \bar{a}_{1} |\psi_1\rangle \otimes |g\rangle\otimes |10
\rangle + \bar{a}_{2} |\psi_2 \rangle \otimes |g\rangle\otimes |01 \rangle
\, ,
\end{equation}
\noindent where $|\bar{a}_1|^2 + |\bar{a}_2|^2 = 1$, $|g\rangle$ corresponds to the atom ground state and $|i j \rangle$, $i, j = \{0, 1\}$ correspond to the number of photons in cavity $A, B$ respectively. The quantum coherence terms between non-exotic trajectories are washed out as expected.
%
However, if the contributions coming from exotic looped trajectories are taken into account, then the full composite system state $|\Psi \rangle $ at the screen immediately before detection would be actually
\begin{eqnarray}\label{eq:St_Sc_ELT}
&|\Psi \rangle = a_1 |\psi_1\rangle \otimes |g\rangle\otimes
|10\rangle + a_2|\psi_2\rangle \otimes |g\rangle\otimes |01 \rangle
\\ \nonumber & {\color{blue}-} a_{12} |\psi_{12}\rangle \otimes |e\rangle\otimes |00
\rangle {\color{blue}-} a_{21}|\psi_{21}\rangle \otimes |e\rangle\otimes |00
\rangle \, ,
\end{eqnarray}
\noindent where $|e \rangle$ corresponds to the atom's excited state and $|a_1|^2+|a_2|^2+|a_{12}|^2+|a_{21}|^2 = 1$. The corresponding reduced density operator for the center of mass subsystem, $\rho _{\mathrm{c.m.}}$ is
\begin{eqnarray}\label{eq:St_Sc_C-of-M}
&\rho _{\mathrm{c.m.}} = |a_1 |^2 |\psi_1 \rangle\langle \psi_1 | +
|a_2 |^2 |\psi_2 \rangle \langle \psi_2 |\\ \nonumber
&+|a_{12} |^2 |\psi_{12} \rangle \langle \psi_{12} |+|a_{21} |^2 |\psi_{21} \rangle \langle \psi_{21} |\\ \nonumber
&+\left[  a_{12} a_{21}^{\ast }\, |\psi_{12} \rangle \langle \psi_{21} | + h.c. \right]\, ,
\end{eqnarray}
\noindent where $h.c.$ is the hermitian conjugate.~This state contain interference terms only between exotic looped trajectories.~Therefore, as stated above, the quantum marking procedure eliminates interferences associated to non-exotic trajectories, but \emph{not} between the exotic looped paths.

\section{Results}
\label{section:results}

\par Now we perform the quantitative analysis of the fringe visibility for exotic trajectories. We adapt the notation used in section \ref{section:dois} to the atom interferometry setting above for the sake of clarity.

\vspace{0.2cm}
\par In order to perform Bell measurements in the superconducting cavity state, we define the Bell basis,

\begin{eqnarray*}
|\psi^+ \rangle &=& \frac{1}{\sqrt{2}} (|00\rangle + |11\rangle)
\end{eqnarray*}
\begin{eqnarray*}
|\psi^- \rangle &=& \frac{1}{\sqrt{2}} (|00\rangle -|11\rangle)
\end{eqnarray*}
\begin{eqnarray*}
|\phi^+ \rangle &=& \frac{1}{\sqrt{2}} (|10\rangle + |01\rangle)
\end{eqnarray*}
\begin{eqnarray*}
|\phi^- \rangle &=& \frac{1}{\sqrt{2}} (|10\rangle -|11\rangle),
\end{eqnarray*}

\noindent which leads to

\begin{eqnarray*}
|00 \rangle &=& \frac{1}{\sqrt{2}} (|\psi^+ \rangle + |\psi^-\rangle)
\end{eqnarray*}
\begin{eqnarray*}
|11 \rangle &=& \frac{1}{\sqrt{2}} (|\psi^+ \rangle - |\psi^-\rangle) \end{eqnarray*}
\begin{eqnarray*}
|10 \rangle &=& \frac{1}{\sqrt{2}} (|\phi^+ \rangle + |\phi^-\rangle) \end{eqnarray*}
\begin{eqnarray*}
|01 \rangle &=& \frac{1}{\sqrt{2}} (|\phi^+ \rangle - |\phi^-\rangle).
\end{eqnarray*}

\noindent Consider the global state (\ref{eq:St_Sc_ELT}) rewritten in the 
Bell basis as
\begin{eqnarray}
&& |\Psi \rangle =  \Big( \frac{a_1}{\sqrt{2}} |\psi_1\rangle + \frac{a_2}{\sqrt{2}} |\psi_2\rangle \Big) \otimes|\phi^+\rangle \otimes |g\rangle + \nonumber \\
&& \Big( \frac{a_1}{\sqrt{2}} |\psi_1\rangle - \frac{a_2}{\sqrt{2}} |\psi_2\rangle \Big) \otimes |\phi^-\rangle \otimes |g\rangle {\color{blue} -} \nonumber \\
&& \Big( \frac{a_{12}}{\sqrt{2}} |\psi_{12}\rangle +
\frac{a_{21}}{\sqrt{2}} |\psi_{21}\rangle \Big) \otimes (|\psi^+
\rangle + |\psi^- \rangle ) \otimes|e\rangle. \nonumber
\end{eqnarray}
\noindent By performing a Bell measurement on the cavity subsystem, there are three possibilities. If one obtains the eigenvalue correspondent to
$|\phi^\pm \rangle$ then the global state collapses to
\begin{eqnarray}
|\Psi\rangle \rightarrow \frac{\hat{P}_{\phi^\pm}
\cdot |\Psi \rangle }{p_{\phi^\pm}},
\end{eqnarray}
\noindent where
\begin{eqnarray}
p_{\phi^\pm} = \text{Tr} [ \hat{P}_{\phi^\pm} |\Psi\rangle \langle \Psi|]
\end{eqnarray}
\noindent and
\begin{eqnarray}
\hspace{-0.5cm} \hat{P}_{\phi^\pm} 
\cdot |\Psi \rangle = \Big( \frac{a_1}{\sqrt{2}} |\psi_1\rangle \pm \frac{a_2}{\sqrt{2}} |\psi_2\rangle \Big) \otimes|\phi^\pm \rangle \otimes |g\rangle \, ,
\end{eqnarray}
\noindent with $\hat{P}_{\phi^\pm}=|\phi^\pm \rangle \langle \phi^\pm|$.~Otherwise $|\Psi \rangle$ collapses to
\begin{eqnarray}
&& |\Psi \rangle \rightarrow |\Psi \rangle_\psi = {\color{blue} - }
\frac{1}{q} \Big(
\frac{a_{12}}{\sqrt{2}} |\psi_{12}\rangle + \frac{a_{21}}{\sqrt{2}}
|\psi_{21}\rangle \Big) \otimes \nonumber \\ && \otimes (|\psi^+
\rangle + |\psi^- \rangle ) \otimes|e\rangle \Big) , \label{NCT}
\end{eqnarray}
\noindent where
\begin{eqnarray}
q &=& \text{Tr} [ \hat{Q} |\Psi \rangle \langle \Psi |], \nonumber \\
\hat{Q} &=& \hat{1} - \hat{P}_{\phi^+} - \hat{P}_{\phi^-}\, .
\end{eqnarray}
\noindent Therefore when joint measurements on the state of the cavities
project such subsystem on $|\phi^+ \rangle$ ($|\phi^- \rangle$), we
observe interference (anti-)fringes stemming from non-exotic paths.~In particular we can obtain exclusively the contribution from exotic
paths to the interference pattern through joint measurements on the
cavities consisting of a superposition of two leading loop
contributions which cross the slit apertures in clockwise and
anticlockwise directions as displayed in equation (\ref{NCT}).

\vspace{0.2cm}
\par Alternatively one may perform measurements on the internal atomic
state. For instance, if the atom is found in the ground state we get
\begin{eqnarray}
|\Psi \rangle \rightarrow  |\Psi \rangle_g  = \frac{(a_1
|\psi_1\rangle |10\rangle + a_2 |\psi_2\rangle |01\rangle) \otimes
|g\rangle}{r} \label{ASMG}
\end{eqnarray}
\noindent with probability $r= \text{Tr} [ |g\rangle \langle g| \cdot |\Psi
\rangle \langle \Psi|]$. 
Notice that in this
case we obtain a pattern clear of exotic paths yet no interference
will be observed as there remains which way information available in
(\ref{ASMG}). In other words the interference between the ordinary
(non-exotic) paths is eliminated due to the entanglement between the
atom's center of mass state and the which-way detector (cavity)
characterized by orthogonal states $|10\rangle$ and $|01\rangle$.
Should the latter be non-orthogonal, the effect would be a decrease
of visibility because entanglement correlations would encode some
which-way knowledge.

\vspace{0.2cm}
\par If otherwise the atom is found in the excited state then, instead of equation (\ref{ASMG}), we will have
\begin{eqnarray}
\Psi \rightarrow |\Psi \rangle_e  = - \frac{(a_{12} |\psi_{12}\rangle
+ a_{21} |\psi_{21}\rangle ) \otimes |00\rangle |e\rangle}{s} \, ,
\label{ASME}
\end{eqnarray}
\noindent with $s= \text{Tr} [|e\rangle \langle e| \cdot |\Psi \rangle \langle\Psi|]$ 
being the probability of measuring
the atom in the excited state. 
Notice that in this case one observes
only exotic paths interference fringes. The interference pattern on
the detection screen is obtained by tracing out the cavity and
atom's internal degrees of freedom, namely $\text{Tr}_{c}$. We
obtain
\begin{eqnarray*}
\mathrm{Tr}_{c}[ |\Psi \rangle_g\, _{g}\langle \Psi |  ] =
\frac{1}{r^2} \Big( |a_1|^2 |\psi_1 \rangle \langle\psi_1| + |a_2|^2
|\psi_2 \rangle \langle\psi_2| \Big)
\end{eqnarray*}
\noindent which has no interference terms, and
\begin{eqnarray}
&& \rho_{\mathrm{et}} \equiv \text{Tr}_{c}[ |\Psi \rangle_e\, _{e}\langle \Psi |  ] = \frac{1}{s^2} \Big[ |a_{12}|^2 |\psi_{12} \rangle \langle\psi_{12}| + \nonumber \\
&& |a_{21}|^2 |\psi_{21} \rangle \langle\psi_{21}| + \Big(a_{12}
a_{21}^*|\psi_{12} \rangle \langle\psi_{21}| + h.c. \Big)\Big],
\label{matden_NC}
\end{eqnarray}
\noindent which has only terms of interference from exotic looped trajectories.
Since the source $S$ is in the axis of symmetry of the experiment (see Figure \ref{fig4}), we can consider without loss of generality that $a_{1}=a_{2}$ and $a_{12}=a_{21}$.~In this way, the state (\ref{matden_NC}) can be rewritten as
\begin{equation}
\rho_{\mathrm{et}}= \mathcal{N}_{\mathrm{et}}^{\,2} \Big[ |\psi_{12} \rangle \langle\psi_{12}| +
|\psi_{21} \rangle \langle\psi_{21}| + \big(|\psi_{12} \rangle \langle\psi_{21}| + h.c. \big)\Big],
\label{matden_NC2}
\end{equation}
with $\mathcal{N}_{\mathrm{et}}=|a_{12}|/s$.~The intensity $I_{\mathrm{et}}(x,t,\tau)$ 
just before being detected in the screen is obtained from (\ref{matden_NC2})
as follows:

\begin{eqnarray}
	I_{\mathrm{et}}(x,t,\tau)&=&\mathcal{N}^{\,2}_{\mathrm{et}}\Bigl\{|\psi_{12}(x,t,\tau)|^2+|\psi_{21}(x,t,\tau)|^2\nonumber\\
	&+&2|\psi_{12}(x,t,\tau)\psi_{21}(x,t,\tau)|\cos(\phi^{21}_{12})
	\Bigr\}
	\label{int_NC}
\end{eqnarray}
\noindent where $\phi^{21}_{12}$ in the cosine denotes the phase difference between $\psi_{12}(x,t,\tau)$ and $\psi_{21}(x,t,\tau)$.
%
%
The wavefunction $\psi_{12}(x,t,\tau)$, and similarly for $\psi_{21}(x,t,\tau)$, is given by \cite{Paz3}
\begin{eqnarray}
\psi _{12}(x,t,\tau)&=&\int_{x_0,x_1,x_2,x_3}
K_{\tau}(x,\tau+\tilde{t};x_{3},\tilde{t})\\
&&\,\times F(x_{3}- d/2)F(x_{2}+d/2)K(1\rightarrow2;2\rightarrow1)\nonumber\\
&&\times\, F(x_{1}- d/2)K_{t}(x_{1},t+\eta;x_{0},0)\psi_{0}(x_{0}),\nonumber
\label{psiNC}
\end{eqnarray}
\noindent where $\tilde{t} = t+2\epsilon$, and where
\begin{equation*}
K(x_{j},t_{j};x_{0},t_0)=\sqrt{\frac{m}{2\pi i\hbar
(t_{j}-t_{0})}}\exp\left[\frac{im(x_{j}-x_{0})^{2}}{2\hbar
(t_{j}-t_0)}\right]
\end{equation*}
\noindent is the free propagator,
\begin{equation*}
F(x_{j})=\exp\left[-\frac{(x_{j})^{2}}{2\beta^{2}}\right]
\end{equation*}
\noindent is the slit transmission function,
\begin{equation*}
\psi_{0}(x_{0})=\frac{1}{\sqrt{\sigma_{0}\sqrt{\pi}}}\exp\left(-\frac{x_{0}^{2}}{2\sigma_{0}^{2}}\right),
\end{equation*}
\noindent is the initial wave packet, and
\bq K(1\rightarrow2;2\rightarrow1)=\sqrt{\frac{m}{4\pi
i\hbar(\epsilon+\eta)}} \times \nonumber \\
\exp\left[\frac{im[(x_{2}-x_{1})^{2}+(x_{3}-x_{2})^{2}]}{4\hbar(\epsilon+\eta)}\right]
\eq
\noindent denotes the free propagator which propagates from slit $1$ to
slit $2$ and from slit $2$ to slit $1$. The parameter $\eta
\rightarrow 0$ is an auxiliary inter slit time parameter, and
$\epsilon$ denotes the time spent from one slit to the next and is
determined by the momentum uncertainty in the $x$-direction, i.e.,
$\epsilon=\frac{d}{\Delta v_{x}}$ ($\Delta v_{x}=\Delta p_{x}/m$),
with $\Delta
p_{x}=\sqrt{\langle\hat{p}^{2}_{x}\rangle-\langle\hat{p}_{x}\rangle^{2}}$,
$\hat{p}_{x}$ being the momentum operator in the $x$-direction. The
time $\epsilon$ is a statistical fluctuation on the time for motion
in the $x$-direction, which has to attain a minimum value $d/\Delta
v_{x}$ in order to guarantee the existence of a exotic trajectory
\cite{Paz4}.

\vspace{0.2cm}
\par After some lengthy algebraic manipulations, we obtain
\begin{eqnarray}\label{psiet12}
\psi_{12}(x,t,\tau)=A_{\mathrm{et}}
\exp\left(-C_{1\mathrm{et}}x^2+C_{2\mathrm{et}}x+C_{3\mathrm{et}}\right)\\
\nonumber \times \exp\left(i\alpha_{\mathrm{et}}
x^2+i\gamma_{\mathrm{et}} x+i
\theta_{\mathrm{et}}+i\mu_{\mathrm{et}}\right).
\end{eqnarray}
\noindent The wave function for the exotic trajectory $21$ (counterclockwise
loop) is obtained by substituting $d$ by $-d$ in Eq.~(\ref{psiet12}), which is given by
\begin{eqnarray}\label{psiet21}
\psi_{21}(x,t,\tau)=A_{\mathrm{et}}
\exp\left(-C_{1\mathrm{et}}x^2-C_{2\mathrm{et}}x+C_{3\mathrm{et}}\right)\nonumber\\
\times \exp\left(i\alpha_{\mathrm{et}} x^2-i\gamma_{\mathrm{et}}
x+i\theta_{\mathrm{et}}+i\mu_{\mathrm{et}}\right).
\end{eqnarray}
\noindent All the coefficients present in equation (\ref{psiet21})
are written out in Appendices 1 and 2 for the sake of completeness. The
indices $R$ and $I$ stand for the real and imaginary part of the
complex numbers that appear in the solutions. As discussed in
\cite{Paz2}, $\mu_{\mathrm{et}}(t,\tau)$ and $\theta_{\mathrm{et}}(t,\tau)$ are phases that do not depend of the transverse position $x$, i.e., they are axial phases. Different from the Gouy phase $\mu _{\mathrm{et}}(t,\tau)$ (App. 2), $\theta_{\mathrm{et}}(t,\tau)$ (App. 1) is a phase that appears as we displace the slit from a given distance away from the origin, which is dependent on the parameter $d$.

\begin{figure}[htp]
	\centering
	\includegraphics[width=5.0 cm]{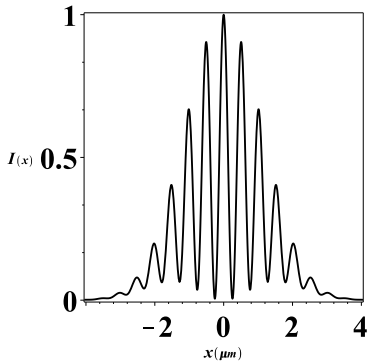}
	\caption{Normalized intensity of only exotic trajectories
		contribution as a function of $x$.} \label{fig5}
\end{figure}

\vspace{0.2cm}
\par The preparation of Rydberg atoms is described in details in \cite{Paulo}. The mass of the Rydberg atoms is
$m=1.44\times10^{-25}\;\mathrm{kg}$, the life time of the excited
state is $30\;\mathrm{ms}$ and the transition frequency from the
ground state to the excited state is $\omega_{ge}=51.099\;\mathrm{GHz}$
which is the same of the fields in cavities $A$ and $B$ (resonant
interaction). The wavelength of the field in these cavities must be
$\lambda={2\pi}c/{\omega}_{ge}$, which have to be microwaves \cite{Raimond}.~We can chose the width of the field and the atom velocity in order to have an interaction time between the atom and cavities of
$t_{i}=1.0\;\mathrm{\mu s}$. For the transverse width of the initial
wavepacket and the double slit apparatus we chose the following
parameters: $\sigma_{0}=10\;\mathrm{nm}$, $\beta=10\;\mathrm{nm}$,
$d=180\;\mathrm{nm}$, $t=20\;\mathrm{\mu s}$, and
$\tau=20\;\mathrm{\mu s}$.~For these parameters we obtain
$\epsilon=3.5\;\mathrm{\mu s}$.~We can observe that the propagation
time $t$, $\epsilon$ and $\tau$ are much smaller than the life time
of the excited state.~This guarantees that the atom does not change
its state in the propagation from the source to the detector by
spontaneous emission.

\vspace{0.2cm}
\par In Figure \ref{fig5} we show the intensity of only the exotic looped paths contribution for Rydberg atoms and the above parameters. Therefore, the results above show that one can prepare a double slit experiment to exhibit the interference of exotic trajectories. Differently from other results that calculate and measure the Sorkin parameter which is produced basically by the interference of exotic and non-exotic trajectories, our results show an interference pattern that can be associated exclusively to the exotic paths. In particular, our results can be important in potential applications involving only such paths.

\section{Discussion}
\label{section:discussion}

\par We studied the double slit experiment with two-level 
atoms and QED cavities positioned in the slit apertures. We prepared
an experimental setup capable to eliminate the contribution to the
interference pattern of non-exotic trajectories.~This is possible by
considering Rydberg atoms interacting resonantly with the QED
cavities. We considered that the source sends the atoms one-by-one in the excited state and that the cavities in the slits are in the fundamental mode. After the atoms pass through one of the cavities it emits a photon, and thus measurement of the cavities' electromagnetic field states enables us to access which-way information about the atom. This procedure eliminates the interference of trajectories that pass through only one slit before arriving at the detector. On the other hand, if the atoms propagate in exotic looped trajectories, the photon emitted in the first passage is reabsorbed in the second, and measurement of the cavities EM field states does not enable us to acquire information about which slit the atom traversed. Therefore, we will observe in the detector an interference pattern exclusive from the exotic looped trajectories. We chose a set of parameters, whose values are from Rubidium atoms and real QED cavities, and we plotted the intensity as a function of $x$ for the interference of exotic trajectories. The approach used in this paper can open an avenue in possible applications where we need to select effects only from the exotic paths' contribution.

\section*{Acknowledgments}
\noindent G.S. and M.S. thank the Departamento de Ci\^{e}ncias
Exatas e Tecnol\'{o}gicas/UESC -- Ilh\'{e}us for the hospitality and
financial support during early stages of the development of this
work. M.S. also acknowledges financial support form the Brazilian
institutions CNPq (Conselho Nacional de Desenvolvimento
Cient\'{i}fico e Tecnol\'{o}gico) and FAPEMIG (Funda\c{c}\~{a}o de
Amparo \`{a} Pesquisa do Estado de Minas Gerais). I. G. da Paz
thanks support from the program PROPESQ (UFPI/PI) under grant number
CCN-007/2016. J.G.O. acknowledges financial support from FAPESB
(Funda\c{c}\~{a}o de Amparo \`{a} Pesquisa do Estado da Bahia),
AUXPE-FAPESB-3336/2014 number 23038.007210/2014-19.

\section*{Appendix 1: Formulae for interference parameters}

\par In the following we present the complete expressions for terms occurring in Eqs. (\ref{psiet12}) and (\ref{psiet21}):

\begin{equation}
A_{\mathrm{et}}=\sqrt{\frac{m^{3}\sqrt{\pi}}{16\hbar^{3}\tau
t\epsilon\sigma_{0}\sqrt{z_{R}^{2}+z_{I}^{2}}}},
\end{equation}

\begin{equation}
C_{1\mathrm{et}}=\frac{m^{2}z_{3R}}{4\hbar^{2}\tau^{2}(z_{3R}^{2}+z_{3I}^{2})},
\end{equation}

\begin{eqnarray}
C_{2\mathrm{et}}&=&-\frac{mdz_{3I}}{4\hbar\tau\beta^{2}(z_{3R}^{2}+z_{3I}^{2})}+\frac{m^{3}dz_{6I}}{64\hbar^{3}\beta^{2}\tau\epsilon^{2}(z_{6R}^{2}+z_{6I}^{2})}
\nonumber
\\&+&\frac{m^2dz_{10R}}{16\hbar^2\tau\beta^{2}(z_{10R}^{2}+z_{10I}^{2})},
\end{eqnarray}

\begin{eqnarray}
C_{3\mathrm{et}}&=&\frac{d^{2}z_{1R}}{16\beta^{4}(z_{1R}^{2}+z_{1I}^{2})}+\frac{d^{2}z_{2R}}{16\beta^{4}\epsilon(z_{2R}^{2}+z_{2I}^{2})}\nonumber
\\&+&\frac{d^{2}z_{3R}}{16\beta^{4}(z_{3R}^{2}+z_{3I}^{2})}-\frac{m^{2}d^{2}z_{4R}}{4^4\beta^{4}\hbar^{2}\epsilon^{2}(z_{4R}^{2}+z_{4I}^{2})}\nonumber
\\&+&\frac{m^4d^{2}z_{5R}}{4^6\hbar^4\epsilon^4\beta^{4}(z_{5R}^{2}+z_{5I}^{2})}-\frac{m^2d^{2}z_{6R}}{2^7\hbar^2\epsilon^2\beta^{4}(z_{6R}^{2}+z_{6I}^{2})}\nonumber
\\&+&\frac{md^{2}z_{7I}}{32\hbar\beta^{4}\epsilon(z_{7R}^{2}+z_{7I}^{2})}-\frac{m^2d^{2}z_{8R}}{4^4\hbar^2\epsilon^2\beta^{4}(z_{8R}^{2}+z_{8I}^{2})}\nonumber
\\&-&\frac{m^3d^{2}z_{9I}}{2^9\hbar^3\epsilon^3\beta^{4}(z_{9R}^{2}+z_{9I}^{2})}+\frac{md^{2}z_{10I}}{32\hbar\epsilon\beta^{4}(z_{10R}^{2}+z_{10I}^{2})}\nonumber
\\ &-&\frac{d^2}{8\beta^2}-\frac{d^2}{4\beta^2},
\end{eqnarray}

\begin{equation}
\alpha_{\mathrm{et}}=\frac{m}{2\hbar\tau}+\frac{m^{2}z_{3I}}{4\hbar^{2}\tau^{2}(z_{3R}^{2}+z_{3I}^{2})},
\end{equation}

\begin{eqnarray}
\gamma_{\mathrm{et}}&=&-\frac{mdz_{3R}}{4\hbar\tau\beta^{2}(z_{3R}^{2}+z_{3I}^{2})}+\frac{m^{3}dz_{6R}}{64\hbar^{3}\beta^{2}\tau\epsilon^{2}(z_{6R}^{2}+z_{6I}^{2})}
\nonumber\\
&-&\frac{m^{2}dz_{10I}}{16\hbar^{2}\tau\epsilon\beta^{2}(z_{10R}^{2}+z_{10I}^{2})},
\end{eqnarray}

\begin{eqnarray}
\theta_{\mathrm{et}}&=&-\frac{d^{2}z_{1I}}{16\beta^{4}(z_{1R}^{2}+z_{1I}^{2})}-\frac{d^{2}z_{2I}}{16\beta^{4}\hbar^{2}\epsilon^{2}(z_{2R}^{2}+z_{2I}^{2})}\nonumber
\\&-&\frac{d^{2}z_{3I}}{16\beta^{4}(z_{3R}^{2}+z_{3I}^{2})}+\frac{m^2d^{2}z_{4I}}{4^4\hbar^2\beta^{4}\epsilon^2(z_{4R}^{2}+z_{4I}^{2})}\nonumber
\\&-&\frac{md^{4}d^2z_{5I}}{4^6\hbar^4\beta^{4}\epsilon^4(z_{5R}^{2}+z_{5I}^{2})}+\frac{m^2d^{2}z_{6I}}{2^7\hbar^2\beta^{4}\epsilon^2(z_{6R}^{2}+z_{6I}^{2})}\nonumber
\\&+&\frac{md^{2}z_{7R}}{32\hbar\beta^{4}\epsilon(z_{7R}^{2}+z_{7I}^{2})}+\frac{m^2d^{2}z_{8I}}{4^2\beta^{4}\epsilon^2(z_{8R}^{2}+z_{8I}^{2})}\nonumber
\\
&-&\frac{m^3d^{2}z_{9R}}{2^9\hbar^3\beta^{4}\epsilon^3(z_{9R}^{2}+z_{9I}^{2})}+\frac{md^{2}z_{10R}}{4^4\hbar\beta^{4}\epsilon(z_{10R}^{2}+z_{10I}^{2})}.
\end{eqnarray}

\vspace{0.3cm}
\section*{Appendix 2: Gouy phase components}

\par In the following we present the full expression of the Gouy phase
for exotic trajectories, i.e.,

\begin{equation}
\mu_{et}(t,\tau)=\frac{1}{2}\arctan\left(\frac{z_{I}}{z_{R}}\right),
\label{ncgouy}
\end{equation}

\noindent where

\begin{eqnarray}
z_{R}&=&(z_{0R}z_{1R}-z_{0I}z_{1I})(z_{2R}z_{3I}+z_{2I}z_{3R})+\nonumber
\\&+&(z_{0R}z_{1I}+z_{0I}z_{1R})(z_{2R}z_{3R}-z_{2I}z_{3I}),
\end{eqnarray}

\noindent and where

\bq
z_{I}&=&(z_{0R}z_{1R}-z_{0I}z_{1I})(z_{2R}z_{3R}-z_{2I}z_{3I})\nonumber\\&-&(z_{0R}z_{1I}+z_{0I}z_{1R})(z_{2R}z_{3I}+z_{2I}z_{3R}).
\eq

\noindent In these expressions, we have:

\begin{equation}
z_{0R}=\frac{1}{2\sigma_{0}^{2}},\;\;z_{0I}=-\frac{m}{2\hbar t},
\end{equation}
\begin{equation}
z_{1R}=\frac{1}{2\beta^{2}}+\frac{m^{2}z_{0R}}{4\hbar^{2}
t^{2}(z_{0R}^{2}+z_{0I}^{2})},\;\;
\end{equation}
\begin{equation}
z_{1I}=-\left(\frac{m}{4\hbar \epsilon}+\frac{m}{2\hbar
t}+\frac{m^{2}z_{0I}}{4\hbar^{2}t^{2}(z_{0R}^{2}+z_{0I}^{2})}\right),
\end{equation}
\begin{equation}
z_{2R}=\frac{1}{2\beta^{2}}+\frac{m^{2}z_{1R}}{16\hbar^{2}\epsilon^{2}(z_{1R}^{2}+z_{1I}^{2})},
\end{equation}
\begin{equation}
z_{2I}=-\left(\frac{m}{2\hbar\epsilon}+\frac{m^{2}z_{1I}}{16\hbar^{2}\epsilon^{2}(z_{1R}^{2}+z_{1I}^{2})}\right),
\end{equation}

\begin{equation}
z_{3R}=\frac{1}{2\beta^{2}}+\frac{m^{2}z_{2R}}{16\hbar^{2}\epsilon^{2}(z_{2R}^{2}+z_{2I}^{2})},
\end{equation}

\begin{equation}
z_{3I}=-\left(\frac{m}{2\hbar
\tau}+\frac{m}{4\hbar\epsilon}+\frac{m^{2}z_{2I}}{16\hbar^{2}\epsilon^{2}(z_{2R}^{2}+z_{2I}^{2})}\right),
\end{equation}

\begin{equation}
z_{4R}=z_{1R}^{2}z_{2R}-z_{1I}^{2}z_{2R}-2z_{1R}z_{1I}z_{2I},
\end{equation}

\begin{equation}
z_{4I}=z_{1R}^{2}z_{2I}-z_{1I}^{2}z_{2I}+2z_{1R}z_{1I}z_{2R},
\end{equation}

\begin{eqnarray}
&&z_{5R}=z_{3R}\big(z_{1R}^{2}z_{2R}^{2}-z_{1R}^{2}z_{2I}^{2}-z_{1I}^{2}z_{2R}^{2}+z_{1I}^{2}z_{2I}^{2}\nonumber
\\ &&-4z_{1R}z_{1I}z_{2R}z_{2I}\big)
-2z_{3I}\big(z_{1R}^{2}z_{2R}z_{2I}-z_{1I}^{2}z_{2R}z_{2I}\nonumber
\\&&+z_{1R}z_{1I}z_{2R}^{2}-z_{1R}z_{1I}z_{2I}^{2}\big),
\end{eqnarray}

\begin{eqnarray}
&&z_{5I}=z_{3I}(z_{1R}^{2}z_{2R}^{2}-z_{1R}^{2}z_{2I}^{2}-z_{1I}^{2}z_{2R}^{2}+z_{1I}^{2}z_{2I}^{2}
\nonumber \\
&&-4z_{1R}z_{1I}z_{2R}z_{2I})+2z_{3R}(z_{1R}^{2}z_{2R}z_{2I}\nonumber
\\&&-z_{1I}^{2}z_{2R}z_{2I}+z_{1R}z_{1I}z_{2R}^{2}-z_{1R}z_{1I}z_{2I}^{2}),
\end{eqnarray}

\begin{equation}
z_{6R}=z_{1R}z_{2R}z_{3R}-z_{1R}z_{2I}z_{3I}-z_{1I}z_{2R}z_{3I}-z_{1I}z_{2I}z_{3R},
\end{equation}

\begin{equation}
z_{6I}=z_{1R}z_{2R}z_{3I}+z_{1R}z_{2I}z_{3R}+z_{1I}z_{2R}z_{3R}-z_{1I}z_{2I}z_{3I},
\end{equation}

\begin{equation}
z_{7R}=z_{1R}z_{2R}-z_{1I}z_{2I},
\end{equation}

\begin{equation}
z_{7I}=z_{1I}z_{2R}+z_{1R}z_{2I},
\end{equation}

\begin{equation}
z_{8R}=(z_{2R}^2-z_{2I}^2)z_{3R}-2z_{2R}z_{2I}z_{3I},
\end{equation}

\begin{equation}
z_{8I}=(z_{2R}^2-z_{2I}^2)z_{3I}+2z_{2R}z_{2I}z_{3R},
\end{equation}

\begin{equation}
z_{9R}=z_{1R}z_{8R}-z_{1I}z_{8I},
\end{equation}

\begin{equation}
z_{9I}=z_{1I}z_{8R}+z_{1R}z_{8I},
\end{equation}

\begin{equation}
z_{10R}=z_{2R}z_{3R}-z_{2I}z_{3I},
\end{equation}

\begin{equation}
z_{10I}=z_{2I}z_{3R}+z_{2R}z_{3I},
\end{equation}



\begin{thebibliography}{99}

\bibitem{Born} M. Born (1926), in J. A. Wheeler and W. H. Zurek, \emph{Quantenmechanik der Sto$\beta$vorg\"ange}, Princeton University Press (1983) 863.

\bibitem{Sorkin} R. D. Sorkin, Mod. Phys. Lett. A 09 (1994) 3119.

\bibitem{Sinha1}
U. Sinha, C. Couteau, T. Jennewein, R. Laflamme, and G. Weihs,
{Science} 329, 418 (2010).

\bibitem{footnote1} On the other hand, paths that go backward in time are also allowed by the path integral formalism. They give rise to the relativistic propagator and one could ask how relativistic corrections to the propagator compare to the Sorkin factor that incorporates exotics paths. This point will be addressed in a forthcoming paper.

\bibitem{Yabuki} H. Yabuki, Int. J. Theor. Phys. 25 (1986) 159.

\bibitem{FeynmanHibbs} R. P. Feynman and A. R. Hibbs,  \textsl{Quantum Mechanics and Path Integrals} (McGraw-Hill, New York, 3rd. ed. 1965).

\bibitem{Raedt}
H. D. Raedt, K. Michielsen, and K. Hess, {Phys. Rev. A} 85,
012101 (2012).

\bibitem{Sinha2}
R. Sawant, J. Samuel, A. Sinha, S. Sinha, and U. Sinha, {Phys. Rev.
    Lett.} 113, 120406 (2014).

\bibitem{Sinha3}
A. Sinha, A. H. Vijay, and U. Sinha, {Scientific Reports}
5, 10304 (2015).


\bibitem{Boyd} O. S. Maga\~{n}a-Loiaza, et. al., Nature Communications, DOI: 10.1038/ncomms13987  (2016).

\bibitem{USinha}
G. Rengaraj, U. Prathwiraj, S. N. Sahoo, R. Somashekhar, and U.
Sinha, \emph{Measuring the deviation from the superposition principle in
interference experiments}, arXiv:1610.09143.

\bibitem{Jin} Fangzhou Jin, et. al., Phys. Rev. A 95 (2017) 012107.

\bibitem{Barnum} H. Barnum, C. M. Lee, C. M. Scandolo and J. H. Selby, \emph{Ruling out higher-order interference from purity principles}, arXiv: 1704.0510.

\bibitem{Lee}
C. M. Lee and J. H. Selby, {New Journ. Phys.} 18, 033023
(2016); C. M. Lee and J. H. Selby, {New Journ. Phys.} 18,
093047 (2016); C. M. Lee and J. H. Selby, {Found. Phys.}
47, 89 (2017).

\bibitem{Park} D.K. Park, O. Moussa, and R. Laflamme, New Journal of Physics 14(11) (2012) 113025.

\bibitem{Niestegge} G. Niestegge, Foundations of Physics 43 (2013) 805.

\bibitem{Paz3} C.H.S. Vieira, H. Alexander, G. de Souza, M. Sampaio, and I. G. da Paz, \emph{Exotic looped trajectories in double-slit experiments with matter waves}, arXiv: 1705.07156.

\bibitem{livro} Serge Haroche, and Jean-Michel Raimond, {\it Exploring the quantum: atoms, cavities, and photons}, Oxford university press, 2006.

\bibitem{atomo} R. G. Hulet, and D. Kleppner, Phys. Rev. Lett. 51, 1430 (1983). 
T. F. Gallagher, {\it Rydberg Atoms}, Cambridge University Press, Cambridge (1994).

\bibitem{interacao} E. T. Jaynes, and F. W. Cummings, Proc. IEEE 51, 89 (1963).

\bibitem{Scully_Englert_Walther} M.O. Scully, B.-G. Englert, and H. Walther, Nature (London) 351 (1991) 111.

\bibitem{Paz4} I. G. da Paz, C. H. S. Vieira, R. Ducharme, L. A. Cabral, H. Alexander, and M. Sampaio, {Phys. Rev. A} 93, 033621 (2016).

\bibitem{Paz2}
C. J. S. Ferreira, L. S. Marinho, T. B. Brasil, L. A. Cabral, J. G.
G. de Oliveira Jr, M. D. R. Sampaio, and I. G. da Paz, {Ann. of Phys.} 362, 473 (2015).

\bibitem{Paulo}
P. Nussenzveig \textit{et. al}. {Phys. Rev. A} 48, 3991
(1993).

\bibitem{Raimond}
J. M. Raimond, M. Brune, and S. Haroche, {Rev. Mod. Phys.}
73, 565 (2001).

\end{thebibliography}
\end{document}